\documentclass[aps,prl,showpacs,amsmath,amssymb,twocolumn,superscriptaddress,floatfix]{revtex4}

\usepackage{graphicx}
\usepackage{dcolumn}
\usepackage{bm}
\usepackage{natbib}
\usepackage{amsmath}

\renewcommand{\vec}[1]{\bm{#1}}
\newcommand{\uvec}[1]{\hat{\vec{#1}}}
\newcommand{\avr}[1]{\left\langle#1\right\rangle}
\newcommand{\Lv}{\mathcal{L}}

\begin{document}
\title{Glass Transition for Driven Granular Fluids}

\author{W. Till Kranz}
\affiliation{Max-Planck-Institut f\"ur Dynamik und Selbstorganisation,
Bunsenstr. 10, 37073 G\"ottingen, Germany}
\affiliation{Georg-August-Universit\"at G\"ottingen, 
Institut f\"ur Theoretische Physik, 
Friedrich-Hund-Platz 1, 37077 G\"ottingen, Germany}
\author{Matthias Sperl}
\affiliation{Institut f\"ur Materialphysik im Weltraum,
Deutsches Zentrum f\"ur Luft- und Raumfahrt, 51170 K\"oln, Germany}
\author{Annette Zippelius}
\affiliation{Georg-August-Universit\"at G\"ottingen, 
Institut f\"ur Theoretische Physik, 
Friedrich-Hund-Platz 1, 37077 G\"ottingen, Germany}
\affiliation{Max-Planck-Institut f\"ur Dynamik und Selbstorganisation,
Bunsenstr. 10, 37073 G\"ottingen, Germany}

\date{\today}
\begin{abstract}

We investigate the dynamics of a driven system of dissipative hard spheres 
in the framework of mode-coupling theory. The dissipation is modeled by 
normal restitution, and driving is applied to individual particles in the 
bulk. In such a system, a glass transition is predicted for a finite 
transition density. For increasing inelasticity, the transition shifts to 
higher densities. Despite the strong driving at high dissipation, the 
transition persists up to the limit of totally inelastic normal 
restitution. 

\end{abstract}


\maketitle

\begin{figure}[htb]
  \includegraphics[width=.4\columnwidth]{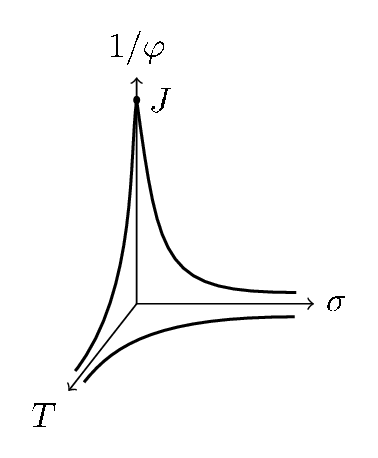}
  \includegraphics{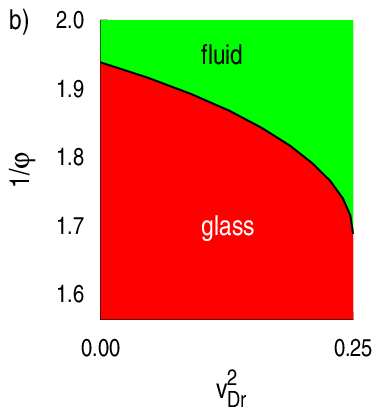}
  \caption{\label{fig:jamming}a) The jamming diagram for sheared
    systems as a function of inverse density $\varphi^{-1}$,
    temperature $T$, and shear stress $\sigma$. b) Jamming diagram for
    driven inelastic hard spheres where the driving power
    $v_{\mathrm{Dr}}^2$ replaces the shear stress.
    In this latter case, the temperature dependence is trivial for $T
    > 0$, and the origin of the graph lies at random-close packing
    (rcp).  }
\end{figure}

The celebrated jamming diagram of Liu and Nagel \cite{Liu1998} generated a 
lot of interest in recent years. The conjecture is that in the space 
spanned by the parameters packing fraction $\varphi$, temperature $T$, and 
external stress $\sigma$, there is a region where the material is solid 
like or jammed, cf. Fig.~\ref{fig:jamming}. This perspective unifies the 
concepts of jamming of macroscopic, athermal particles and of the glass 
transition of microscopic, thermal particles. A lot of work has been 
devoted to the point J --- the arrest of static or quasi static granular 
assemblies \cite{OHern2003}. Similarly, the glass transition of a 
supercooled molecular liquid has been studied extensively, corresponding 
to a transition line in the $\varphi$-$T$-plane, which in the case of hard 
spheres is parallel to the $T$-axis. In such a system of elastic hard 
spheres, the glass transition is described well by the mode-coupling 
theory (MCT) \cite{Bengtzelius1984,Megen1991}. Whether or not the unified 
picture of the jamming diagram holds is still a matter of debate, though.

It was shown by theory and computer simulation that both Newtonian and
Brownian equations of motion yield the same glassy dynamics
\cite{Szamel1991,Goetze2009,Gleim1998}.  The situation is different
when the system is subject to shear. MCT has been extended recently to
the case of colloidal suspensions under shear \cite{Fuchs2002c}, and
for this case it was shown that any finite shear rate is able to
destroy the glass transition. While remnants of the glass transition
are still affecting the dynamics, full arrest is no longer
possible. Another scenario is proposed by the above mentioned jamming
diagram where applied shear stress unjams the system but can be
compensated by higher density or lower temperature,
cf. Fig.~\ref{fig:jamming}(a).

It is the objective of the present paper to investigate the possibility of 
a glass transition in a granular fluid in a steady state when dissipation 
is balanced by bulk driving. Experiments by Abate and Durian 
\cite{Abate2006} and Reis \textit{et al.} \cite{Reis2007} showed 
indications of a granular glass transition in such fluidized granular 
systems in two dimensions. These observations would fit into a modified 
jamming diagram, where the shear-stress axis $\sigma$ is replaced by an 
axis that quantifies the driving force $v_\mathrm{Dr}$ used to compensate 
for the dissipative interactions among the granular particles. In the 
following, we shall investigate if a granular glass transition can exist 
and how such a transition can be described by an appropriate theory.
It will be demonstrated that (1) the combination of MCT \cite{Goetze2009} 
with granular kinetic theory \cite{Aspelmeier2001} predicts a glass 
transition for a driven dissipative system, (2) the nature of the 
transition depends on the degree of dissipation, and (3) granular dynamics 
cannot be scaled onto Brownian or Newtonian dynamics.

We consider the non-equilibrium stationary state of a driven granular
fluid comprised of $N\to\infty$ identical hard spheres of diameter $d$
and mass $m$ in a volume $V$. Energy dissipation in binary collisions
is modeled by a coefficient of normal restitution $\varepsilon$. Due to
the energy loss in the collisions the system needs to be driven in
order to achieve a stationary state. We use a simple bulk driving
mechanism, e.g., as in air fluidized beds \cite{Abate2006}. The particles 
are kicked with frequency $\omega_\mathrm{Dr}$ according to
\begin{equation}\label{eq:drive}
\mathbf{v}_i^\prime (t) = \mathbf{v}_i(t) + 
v_\mathrm{Dr} \bm{\xi}_i(t).\notag
\end{equation}
The driving amplitude, $v_\mathrm{Dr}$, is constant and the direction
of the kick, $\bm\xi_i(t)$, is chosen randomly from a Gaussian
distribution, $P(\bm\xi)$, with unit variance. To ensure momentum
conservation, we choose pairs of neighboring particles and kick them
in opposite directions~\cite{Fiege2009}. A stationary state is
reached, when the energy loss due to collisions is balanced by the
energy input due to driving. A simple estimate relates the properties
of the driving ($\omega_{\mathrm{Dr}},v_{\mathrm{Dr}}$) to the
collision frequency $\omega_{\mathrm{coll}}$:
\begin{equation}\label{eq:balance}
\omega_\mathrm{coll}(1-\varepsilon^2) \frac{T}{4}
= \omega_\mathrm{Dr} v_\mathrm{Dr}^2\,.\notag
\end{equation}
The time evolution of the system consists of ballistic motion in between 
binary collisions and random kicks. These can be formally incorporated in 
a Pseudo-Liouville operator $\Lv_{+}$ \cite{Huthmann1997,Aspelmeier2001}, 
which generates the time evolution of an observable, such as the density
\begin{equation}\label{eq:pLiou}
  \rho_q(t) = \frac1N\sum_i\exp(i\vec q\cdot\vec r_i(t)) =
  \exp(it\Lv_{+})\rho_q(0)\,.\notag
\end{equation}
For the discussion of the long-time dynamics, the central quantity of
interest is the density correlation function,
\begin{equation}\label{eq:corr}
  F(q, t) = \!\int\! d^3\xi P(\bm\xi)\!
  \int\! d\Gamma w(\Gamma)\rho^*_q(0)\rho_q(t)
  =:\avr{\rho_q(0)|\rho_q(t)}\,,\notag
\end{equation}
which is directly accessible from computer simulations and experiments. 
Here, $w(\Gamma)$ is the stationary $N$-particle distribution and 
$\Gamma=\{\vec {r}_i, \vec v_i\}_{i=1}^N$ denotes a point in phase space.

Following Mori and Zwanzig, the normalized correlation function, 
$\phi_q(t) = F(q,t)/S(q) = F(q,t)/F(q, 0)$, can be represented in terms of 
restoring forces and a memory kernel
\cite{Hansen1986},
\begin{equation}
  \label{eq:eom}
  \left(\partial_t^2 + \nu_q\partial_t + \Omega^2_q\right)\phi_q(t)
  = -\!\int_0^t\!d\tau M_q(t-\tau)\partial_{\tau}\phi_q(\tau),
\end{equation}
with
\begin{equation}
  \nu_q = \frac{N}{v_0^2}\avr{j_q^L|\Lv_+j_q^L}, \quad
  \Omega_q^2 = \frac{N^2}{v_0^2S_q}\,
  \avr{\rho_q|\Lv_+j_q^L}\avr{j_q^L|\Lv_+\rho_q}\,.\notag
\end{equation}
The thermal velocity $v_0 = \sqrt{T/m}$ and initial conditions are 
$\phi_q(0) = 1, \partial_t\phi_q(0) = 0$.

Detailed balance does not hold and consequently, the transition rates of 
forward and backward reactions are not simply related, 
$\avr{\rho_q|\Lv_+j_q^L}\ne\avr{j_q^L|\Lv_+\rho_q}^*$, as it would be the 
case in equilibrium systems. For the same reason, the memory kernel, 
$M_q(t) = \avr{R_q^{\dagger}|R_q(t)}$, is now given by the 
\emph{cross}-correlation of two unequal fluctuating forces, $R_q\ne 
R_q^{\dagger}$, driven by the reduced dynamics. Details of the calculation 
can be found elsewhere \cite{Kranz2010}. The representation in 
Eq.~(\ref{eq:eom}) is exact. It correctly accounts for the conservation 
laws for the density, $\rho_q(t)$ and the longitudinal momentum, $qj_L(t) 
= \partial_t\rho_q(t)$. Energy is not conserved in a granular medium and 
transverse momentum is decoupled. Hence the representation guarantees the 
correct hydrodynamic limit of $\phi_q(t)$.

To proceed, we have to resort to approximations. First, we need an
approximate form of the $N$-particle distribution to compute static
(equal time) correlations.  We assume that positions and velocities
are uncorrelated $w(\Gamma) = w_r(\{\vec r_i\})w_v(\{\vec v_i\})$ and
that the velocity distribution factorizes into a product of
one-particle distributions, $w_v(\{\vec v_i\}) = \prod_iw_1(\vec
v_i)$. The precise from of $w_1(\vec v)$ is not needed, it only has to
satisfy $\avr{\vec v} = 0$ and $\avr{\vec v^2} = 3T/m <
\infty$. Furthermore, the system is assumed to be isotropic and
homogeneous except for the excluded volume: $w_r(\{\vec
r_i\})=\prod_{i<j}\theta(r_{ij}-d)$. For the restoring forces we find
\begin{equation}
  \label{eq:3}
  \begin{gathered}
    \nu_q = -i\omega_E\frac{1+\varepsilon}{2}[1 - j_0(qd) + 2j_2(qd)]\,,\\
    \Omega_q^2 = \frac{q^2v_0^2}{S_q}\left( 
      \frac{1+\varepsilon}{2} + \frac{1-\varepsilon}{2}S_q 
    \right)\,,
  \end{gathered}\notag
\end{equation}
with $\omega_E$ the Enskog frequency for the elastic case.

An additional approximation is necessary to compute the memory kernel.
The success of MCT for the description of dense molecular and colloidal 
fluids motivates its application also to dissipative granular fluids. 
First, we project the fluctuating forces onto products of densities. 
Second, the resulting higher order correlations are factorized into pair
correlations. Thereby, one generates an explicit expression for
$M_q(t) = m_q(t)\Omega_q^2$ in terms of $\phi_q(t)$ \cite{Kranz2010},
\begin{equation}
  \label{eq:mem}
  m_q[\phi](t) \approx A_q(\varepsilon)\frac{nS_q}{q^2}
  \mspace{-4mu}\int d^3k\;V_{\vec q\vec k}\phi_{\vec k}(t) \phi_{\vec q - 
\vec k}(t) 
\end{equation}
with $V_{\vec q\vec k}$ given by
\begin{equation} 
  \label{eq:mem:vertex} 
  V_{\vec q\vec k} = S_{\vec k} S_{\vec q -\vec k} 
  \left[\uvec q\cdot\vec k \,c_{\vec k} + \uvec 
    q\cdot(\vec q - \vec k)\,c_{\vec q -\vec k}\right]^2\notag
\end{equation} 
The direct correlation function, $c_q$, is simply related to the
static structure factor via the Ornstein-Zernike equation $nc_q \equiv
1 - S_q^{-1}$ and $A_q(\varepsilon) = [1 + (1 - \varepsilon)S_q/(1 +
\varepsilon)]^{-1}$ depends on $\varepsilon$ explicitly.
Inserting the mode coupling approximation, Eq.~\eqref{eq:mem}, into 
Eq.~\eqref{eq:eom}, we get a self-consistency equation for the scattering 
function $\phi_q(t)$. The only further input that is required is the 
static structure factor $S_q$. For simplicity, we use the elastic 
Percus-Yevick expression \cite{Hansen1986} here. Future work 
\cite{Kranz2010} will study the influence of a more precise structure 
factor that depends on the coefficient of restitution $\varepsilon$.

A memory function under driving is not necessarily positive and might not 
even be a real function \cite{Gazuz2009}. Hence, it is surprising that for 
a driven granular fluid the only change compared to the elastic case is 
the $\varepsilon$-dependent prefactor $A_q(\varepsilon) > 0$, see 
Eq.~(\ref{eq:mem}). Consequently, the memory kernel itself is positive and 
a vast amount of work that has been devoted to the mathematical structure 
of the standard mode coupling equations (for a compilation see, e.g., 
\cite{Goetze2009}) is readily applicable to the granular system. This 
finding is quite remarkable because it implies that a large number of 
results derived for equilibrium systems also holds for a system far from 
equilibrium. In particular, a positive memory function in 
Eq.~(\ref{eq:mem}) guarantees positive spectra. On the other hand, the 
well established universality of glassy dynamics is broken for granular 
fluids since the memory kernel depends on $\varepsilon$ explicitly.

\begin{figure}[htb]
\includegraphics[width=\columnwidth]{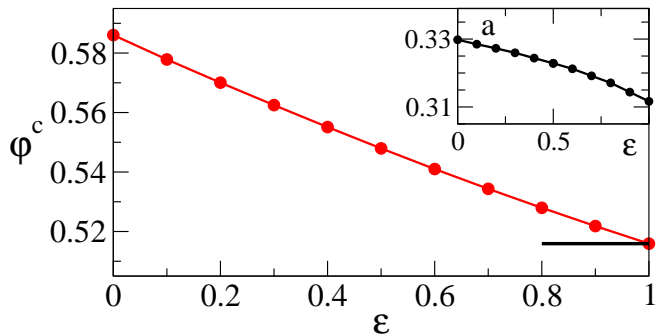}
\caption{\label{fig:phic}Transition density $\varphi^c$ as a function
of the coefficient of restitution $\varepsilon$. The short horizontal
bar indicates the result for the elastic case for $\varepsilon = 1$.
The inset shows the evolution of the critical exponent $a$ with 
$\varepsilon$.
}
\end{figure}

A glass transition is signaled by the appearance of a time persistent part 
of the density correlations $f_q := \lim_{t \to \infty}\phi_q(t)$ in 
Eq.~(\ref{eq:eom}). In this limit, Eq.~(\ref{eq:eom}) reduces to the 
algebraic equation $f_q/(1-f_q) = m_q[f]$ which is solved readily by 
standard procedures \cite{Franosch1997}. For all values of the coefficient 
of restitution, $0\leq\varepsilon\leq1$, an ideal glass transition of the 
driven granular fluid is found with a transition density 
$\varphi^c(\varepsilon)$, cf. Fig.~\ref{fig:phic}. For increasing 
dissipation, i.e., smaller $\varepsilon$, the glass transition is shifted 
to higher densities. This can be understood as follows: For increased 
dissipation, $\varepsilon < 1$, the prefactor $A_q(\varepsilon)$ in 
Eq.~(\ref{eq:mem}) becomes smaller than unity. Dissipation and driving 
hence weaken the memory effects and destabilize the glass. This needs to 
be compensated by a higher density. The resulting compensation of enhanced 
dissipation and driving by increased density can be represented in a 
corresponding jamming diagram as shown in Fig.~\ref{fig:jamming}b, where 
for simplicity the mean field relation $v_\mathrm{Dr}^2 \propto (1 - 
\varepsilon)^2/4$ was used. It might be reassuring that, although the 
critical density increases with increasing dissipation, it stays below the 
density of random-close-packing.

\begin{figure}[htb]
\includegraphics[width=\columnwidth]{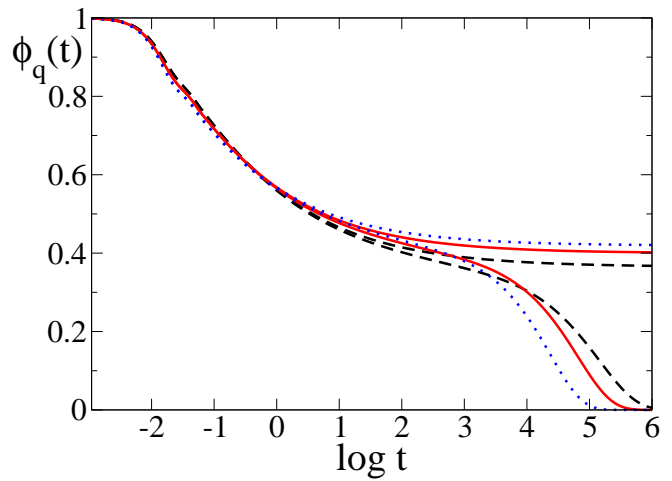}
\caption{\label{fig:phiqt}Dynamics of the coherent density correlator 
$\phi_q(t)$ for the wave vector $qd = 4.2$ at the respective critical 
densities $\varphi^c(\varepsilon)$ and at $\varphi = 0.999 
\varphi^c(\varepsilon)$ for $\varepsilon=1.0$ (dashed lines, elastic 
case), $0.5$ (full lines), and $0.0$ (dotted lines).}
\end{figure}

The full dynamics of Eq.~\eqref{eq:eom} is shown in Fig.~\ref{fig:phiqt}.  
The time dependent scattering function $\phi_q(t)$ with a density close to 
the glass transition density $\varphi_c$ shows the generic two step 
relaxation. After an initial fast relaxation, the scattering function 
approaches a plateau $\phi_q(t) \simeq f_q$ and decays to zero only for 
very long times, provided the density is still below the critical one. 
Close to the transition point a critical decay is found onto these 
plateaus, $\phi_q(t) - f_q \propto t^{-a}$. The variation of the critical 
exponent $a$ is shown in the inset of Fig.~\ref{fig:phic}. For values 
below the transition, $\varphi < \varphi^c$, a second power law describes 
the decay from the plateau, known as the von-Schweidler law, $\phi_q(t) - 
f_q \propto t^{b}$. The exponent $b$ (not shown here) also varies with the 
coefficient of restitution $\varepsilon$, and is uniquely related to $a$.

\begin{figure}[htb]
\includegraphics[width=\columnwidth]{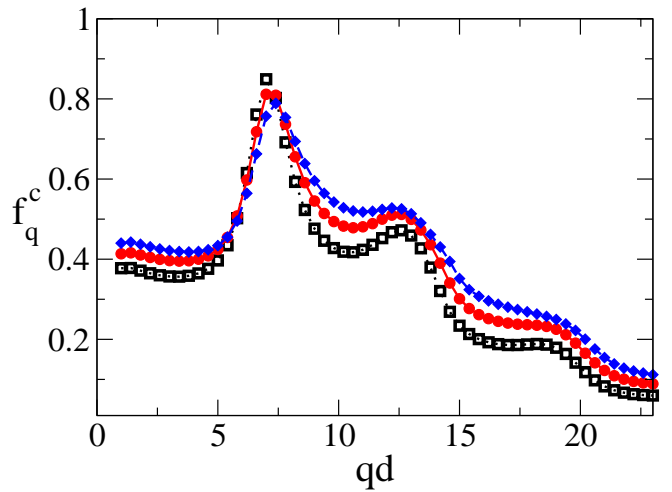}
\caption{\label{fig:fqeps}Critical glass form-factors $f_q^c$ for 
coefficient of restitution $\varepsilon = 1.0$ (empty squares, elastic 
case), 0.5 (filled circles), and 0.0 (filled diamonds) as a function of 
wave number $qd$.}
\end{figure}

At the transition point and beyond, $\varphi > \varphi^c$, the correlation 
function assumes a finite long-time limit $f_q > 0$ which sets in at a 
critical plateau value $f_q^c$. These values are shown in 
Fig.~\ref{fig:fqeps} for various values of $\varepsilon$. The increasing 
dissipation has three noticeable effects: (1) correlations at small 
wavenumbers are enhanced, (2) oscillations reflecting the local structure 
become less pronounced, and (3) the localization length (indicated by the 
inverse of the width of the $f_q$ distribution) decreases. The last 
finding is a consequence of the glass transition taking place at a higher 
density, cf. Fig.~\ref{fig:phic}.

\begin{figure}[hbt]
\includegraphics[width=\columnwidth]{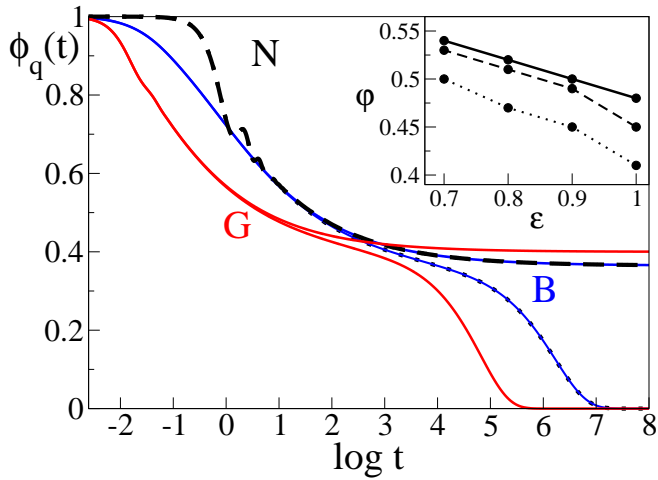}
\caption{\label{fig:nbg}Dynamic scattering function $\phi_q(t)$ for $qd = 
4.2$ at the glass transition, $\varphi = \varphi_c$, and for slightly 
lower volume fraction, $\varphi = 0.999\varphi_c$ for Newtonian dynamics 
(N) with $\nu_q = 0$, Brownian dynamics (B), and granular dynamics (G) 
with coefficient of restitution $\varepsilon=0.5$. The Newtonian dynamics 
is scaled along the time axis to match the Brownian dynamics at long 
times.  No such rescaling is possible for the granular dynamics. The inset 
shows three lines of equal diffusivity (from top to bottom: $D=0.7$, 0.8, 
and 0.9 in relative units) taken from the data of a computer simulation 
\cite{Fiege2009}.
}
\end{figure}

It was shown earlier that Newtonian (N) and Brownian (B) systems show the 
same glassy dynamics \cite{Gleim1998}. This is shown for the MCT dynamics 
in Fig.~\ref{fig:nbg} for curves N and B, where N only needed to be 
shifted along the time axis to match B. In contrast, due to the explicit 
dependence on $\varepsilon$ in Eq.~\eqref{eq:mem}, for any $\varepsilon < 
1$, the granular long-time dynamics (G) cannot be scaled on top of the 
Newtonian or Brownian results. In addition, for granular dynamics at 
different $\varepsilon$ there also exists no single master curve, cf. 
Fig.~\ref{fig:phiqt}. Therefore, granular dynamics leads to a 
fundamentally different long-time behavior.

In conclusion, we have shown that MCT can be extended to granular fluids, 
which are in a steady state far from equilibrium. It can be shown that the 
resulting memory kernel in Eq.~(\ref{eq:mem}) is positive, and this allows 
that most mathematical theorems from equilibrium MCT carry over to the 
driven granular case. An ideal glass transition is observed for all values 
of the dissipation or equivalently all values of the driving. Since the 
balance between dissipation and driving implies that the driving amplitude 
is proportional to the inelasticity $v_\mathrm{Dr}^2 \propto 
(1-\varepsilon)^2$, the glass transition defines a line in the 
$\varphi^{-1}$-versus-$v_{\mathrm{Dr}}^2$ plane of a generalized jamming 
diagram, cf. Fig.~\ref{fig:jamming}. The universality known for Newtonian 
and Brownian dynamics is broken in the granular case -- however, the 
predicted differences in the critical exponents, cf. inset of 
Fig.~\ref{fig:phic}, and glass form factors, cf. Fig.~\ref{fig:fqeps}, are 
relatively small. Comparably large changes with increased dissipation are 
expected in the transition densities as shown in Fig.~\ref{fig:phic}. This 
prediction can be supported by looking at a precursor of the 
glass-transition line in computer-simulation data: For increasing 
dissipation, points of equal diffusivity are found to be shifted to higher 
densities (inset of Fig.~\ref{fig:nbg}) in accordance with our predictions 
for the glass transition in Fig.~\ref{fig:phic}. Therefore, we expect our 
results to be testable in further computer simulation studies and also in 
experiments.

\begin{acknowledgments}
We thank T. Aspelmeier and A. Fiege for interesting dicussions.
This work was supported by DFG Sp714/3-1 and BMWi~50WM0741.
\end{acknowledgments}

\bibliographystyle{apsrev}
\bibliography{lit,add}

\end{document}